\documentclass[galaxies,article,accept,moreauthors,dvipdfm,12pt,a4paper]{mdpi}
\lastpage{x}
\doinum{10.3390/------}
\pubvolume{xx}
\pubyear{2015}
\history{Received: xx / Accepted: xx / Published: xx}

\usepackage{subfigure,epsfig}
\usepackage{natbib}
\Title{Back to the green valley: how to rejuvenate an S0 galaxy through minor mergers}

\Author{Michela Mapelli$^{1}$*} 

\address{%
$^{1}$ INAF -- Astronomical Observatory of Padova, Vicolo dell'Osservatorio 5, 35122, Padova, Italy} 

\corres{michela.mapelli@oapd.inaf.it}

\abstract{About half of the S0 galaxies in the nearby Universe show signatures of recent or ongoing star formation. Whether these S0 galaxies were rejuvenated by the accretion of fresh gas is still controversial. We study minor mergers of a gas-rich dwarf galaxy with an S0 galaxy, by means of N-body smoothed-particle hydrodynamics simulations. We find that minor mergers trigger episodes of star formation in the S0 galaxy, lasting for $\sim{}10$ Gyr. One of the most important fingerprints of the merger is the formation of a gas ring in the S0 galaxy. The ring  is reminiscent of the orbit of the satellite galaxy, and its lifetime depends on the merger properties: polar and counter-rotating satellite galaxies induce the formation of long-lived smooth gas rings.}

\keyword{Galaxies: interactions; Methods: numerical; Galaxies: kinematics and dynamics; Galaxies: ISM; galaxies: star formation}

\begin{document}


\section{Introduction}

S0 galaxies are intriguing objects:  they are early-type galaxies (ETGs), but are characterized by a stellar disc (although with no spiral arms). About half of S0 galaxies \cite{Salim2012} lie in the region of the colour-magnitude diagram (CMD) of galaxies that is called {\it green valley}. With respect to the  blue cloud (where galaxies have blue colours and are actively star forming) and to the  red sequence (i.e. the sequence of red and nearly passive galaxies),  galaxies that populate the green valley are red but `not so dead', i.e. they still form stars at some level. 

\begin{figure*}
\center{{
\epsfig{figure=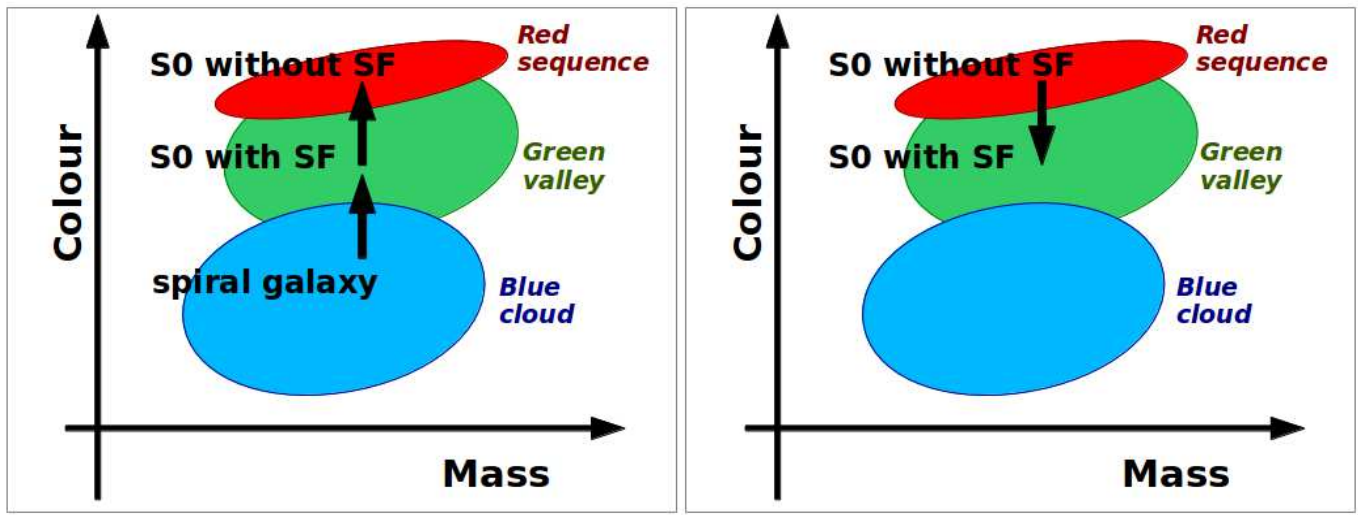,width=18.0cm} 
}}
\caption{\label{fig:fig1}
Cartoon of the colour-magnitude diagram (CMD) of galaxies. The blue, green and red ellipses indicate the blue cloud, the green valley and the red sequence, respectively. Left-hand panel: scenario of star formation (SF) fading (e.g. \cite{Salim2012}). An S0 galaxy enters the green valley during its transformation from a star-forming galaxy to a passive galaxy. Right-hand panel: rejuvenation scenario. An S0 galaxy enters the green valley thanks to the accretion of fresh gas, which re-activates star formation.
}
\end{figure*}

Where do S0 galaxies that lie in the green valley come from? A possible explanation is that S0 galaxies cross the green valley during their transformation from actively star forming to passive galaxies. According to an alternative scenario, star-forming S0 galaxies are `rejuvenated' ETGs: after having already reached the red sequence, they came back to the green valley, because star formation was re-activated thanks to the accretion of fresh gas (Fig.~\ref{fig:fig1}). 

Moreover, most of star-forming S0 galaxies are characterized by the presence of rings of gas and/or young stars \cite{Thilker2007, Aguerri2009}.  There are at least three kinds of such rings: (i) nuclear rings (surrounding the nucleus of the galaxy), (ii) inner rings (generally surrounding the bar, if present), and (iii) outer rings (more external than the former ones), including polar rings. 

What drives the formation of these rings is controversial. It might be connected with resonances induced by the rotation of the bar (i.e. either an inner Linblad's resonance, or an outer Linblad's resonance, or a ultraharmonic resonance, \cite{Binney1987, butacombes1996}). This is supported by the high fraction of S0 galaxies hosting both a ring and a bar (e.g.~\cite{Laurikainen2013}). However, a resonance cannot be invoked for the outer polar rings (e.g. \cite{Iodice2006,Marino2009,Moiseev2011}), as well as for the most asymmetric and off-centre rings (e.g. \cite{Thilker2010}). 

Alternative scenarios to a bar-driven resonance include gas-rich minor or major mergers, and accretion from filaments/clouds of intergalactic medium (e.g. \cite{Bekki1997,Struck1999,Eliche2011}).
Evidences of gas accretion either from filaments or from interactions with gas-rich satellite galaxies have been found in a large number of galaxies (\cite{Sancisi2008} for a review). 
\cite{Marino2011} study five barred S0 galaxies with outer rings in UV. For at least three of these galaxies (NGC~1533, NGC~2962 and NGC~2974), several hints suggest recent accretion of gas from satellite galaxies. Thus, \cite{Marino2011} hypothesize that the outer rings in their sample are an effect of (bar-driven) secular evolution, but at the same time this process was aided by fresh gas accreted from gas-rich satellites.

 The major critique to the minor merger scenario is that star formation episodes triggered by minor mergers are (too) short lived to explain the majority of rejuvenated S0 galaxies. Moreover, rings that form from minor mergers are expected to be irregular and clumpy structures, rather than smooth and regular rings (like those that are observed in the majority of rejuvenated S0 galaxies \cite{Salim2012}).

 This proceeding summarizes the results published in \cite{Mapelli2015}. By means of N-body/smoothed-particle hydrodynamics (SPH) simulations (described in Section~2), we show that minor mergers with gas-rich satellite galaxies can efficiently re-activate star formation in S0 galaxies (Section~3.1). The burst of star formation induced by a minor merger can be extremely long-lived ($\sim{}10$ Gyr). Our simulations show that ring-like structures form as a consequence of gas stripping during the merger process. Polar rings and counter-rotating coplanar rings are more long-lived than similar corotating structures. Long-lived rings appear as smooth and regular structures (Section~3.2).


\section{Numerical methods}

We simulate close encounters (flybys) between a primary S0 galaxy and a secondary gas-rich galaxy. The initial conditions for both the primary galaxy and the secondary galaxy in the $N-$body model are generated by using an upgraded version of the code described in \cite{Widrow2008}. The code generates self-consistent disc-bulge-halo galaxy models, derived from explicit distribution functions for each component, which are very close to equilibrium. In particular, the halo is modelled with a Navarro, Frenk \&{} White (1996, NFW \cite{Navarro1996}) profile. We use an exponential disc model \cite{Hernquist1993}, while the bulge is spherical and comes from a generalization of the Sersic law \cite{Widrow2008}. Both the primary and the secondary galaxy have a stellar bulge and a stellar disc. The total mass of the secondary is $\sim{}1/20$ of the mass of the primary. 
 The giant S0 galaxy has no gas, whereas the secondary galaxy has an initial gas mass of $1.38\times{}10^{8}$ M$_\odot{}$, distributed according to an exponential disc. The assumption that the primary has no gas is quite unrealistic, but allows to distinguish between the triggering of star formation in the pre-existing gaseous disc of the S0 galaxy and the effect of fresh gas accretion from the companion galaxy. 

The adopted orbits are nearly parabolic (eccentricity $e=1.003$, specific orbital energy $E_s=3.8\times{}10^3$ km$^2$ s$^{-2}$, specific orbital angular momentum $L_s= 2.0\times{}10^3$ km s$^{-1}$ kpc, impact parameter $b=10$ kpc), in agreement with predictions from cosmological simulations. More details on the simulations can be found in Table~1 and in \cite{Mapelli2015}.

\begin{table}
\begin{center}
\caption{Initial conditions of the $N-$body simulations: masses and scale lengths.}
 \leavevmode
\begin{tabular}[!h]{lll}
\hline
Model galaxy properties & Primary & Secondary \\
\hline
$M_{\rm DM}$ ($10^{11}$ M$_\odot{}$)          & 7.0  & 0.3\\
$M_\ast{}$ ($10^{10}$ M$_\odot{}$)      & 7.0   & 0.2\\
$f_{\rm b/d}$                               &  0.25 & 0.25 \\
$M_{\rm G}$ ($10^{8}$ M$_\odot{}$)    & 0 &  1.38 \\
$R_{\rm s}$ (kpc) & 6.0 & 3.0\\
$R_{\rm d}$ (kpc) & 3.7 & 3.0 \\
$h_{\rm d}$ (kpc) & 0.37 & 0.30 \\
$r_{\rm B}$ (kpc) & 0.6 & 0.6 \\
\noalign{\vspace{0.1cm}}
\hline
\end{tabular}
\begin{flushleft}
\footnotesize{$M_{\rm DM}$ is the dark matter mass; $M_\ast{}$ is the total stellar mass of the galaxy (including both bulge and disc); $f_{\rm b/d}$  is the bulge-to-disc mass ratio; $M_{\rm G}$ is the total gas mass. The primary has no gas, while the gas of the secondary is distributed according to an exponential disc \cite{Hernquist1993}, with the same parameters (scale length and height) as the stellar disc. $R_{\rm s}$ is the NFW scale radius $R_{\rm s}\equiv{}R_{200}/c$, where $R_{200}$ is the virial radius of the halo \cite{Navarro1996} and $c$ the concentration (here we assume $c=12$ for both galaxies). $R_{\rm d}$ and $h_{\rm d}$ are the disc scale length and height, respectively \cite{Hernquist1993}, while $r_{\rm B}$ is  the bulge exponential scale-length (according to  Prugniel and Simien's deprojection of the Sersic profile, \cite{Widrow2008}).}
\end{flushleft}
\end{center}
\end{table}

In this  proceeding, we present three different simulations. 
In run~1, the secondary galaxy is coplanar with the main galaxy, and its orbital angular momentum has an opposite direction with respect to the spin of the main galaxy (counter-rotating coplanar orbit). In run~2, the secondary galaxy is coplanar with the main galaxy, and its orbital angular momentum has the same direction as the spin of the main galaxy (corotating coplanar orbit). Finally, in run~3, the orbit of the secondary galaxy is polar with respect to the disc of the main galaxy. Runs~1, 2 and 3 correspond to runs~A, B and D in \cite{Mapelli2015}.

In all the simulations, the particle mass in the primary galaxy is $2.5\times{}10^5$ M$_\odot{}$ and $5\times{}10^4$ M$_\odot{}$ for dark matter  and stars, respectively. The particle mass in the secondary galaxy is  $2.5\times{}10^4$ M$_\odot{}$ for dark matter and $5\times{}10^3$ M$_\odot{}$ for both stars and gas. 
The softening length is $\epsilon{}=100$ pc.

We simulate the evolution of the models with the $N-$body/SPH tree code {\sc gasoline} \cite{Wadsley2004}. Radiative cooling, star formation and supernova blastwave feedback are enabled, as described in \cite{Stinson2009}. The adopted parameters for star formation and feedback  are the same as used simulations of galaxy-galaxy collisions \cite{MapelliMayer2012,Mapelli2012,Mapelli2013a,Mapelli2013b}. 

\section{Results and Discussion}
\subsection{Reactivating star formation}

\begin{figure*}
\center{{
\epsfig{figure=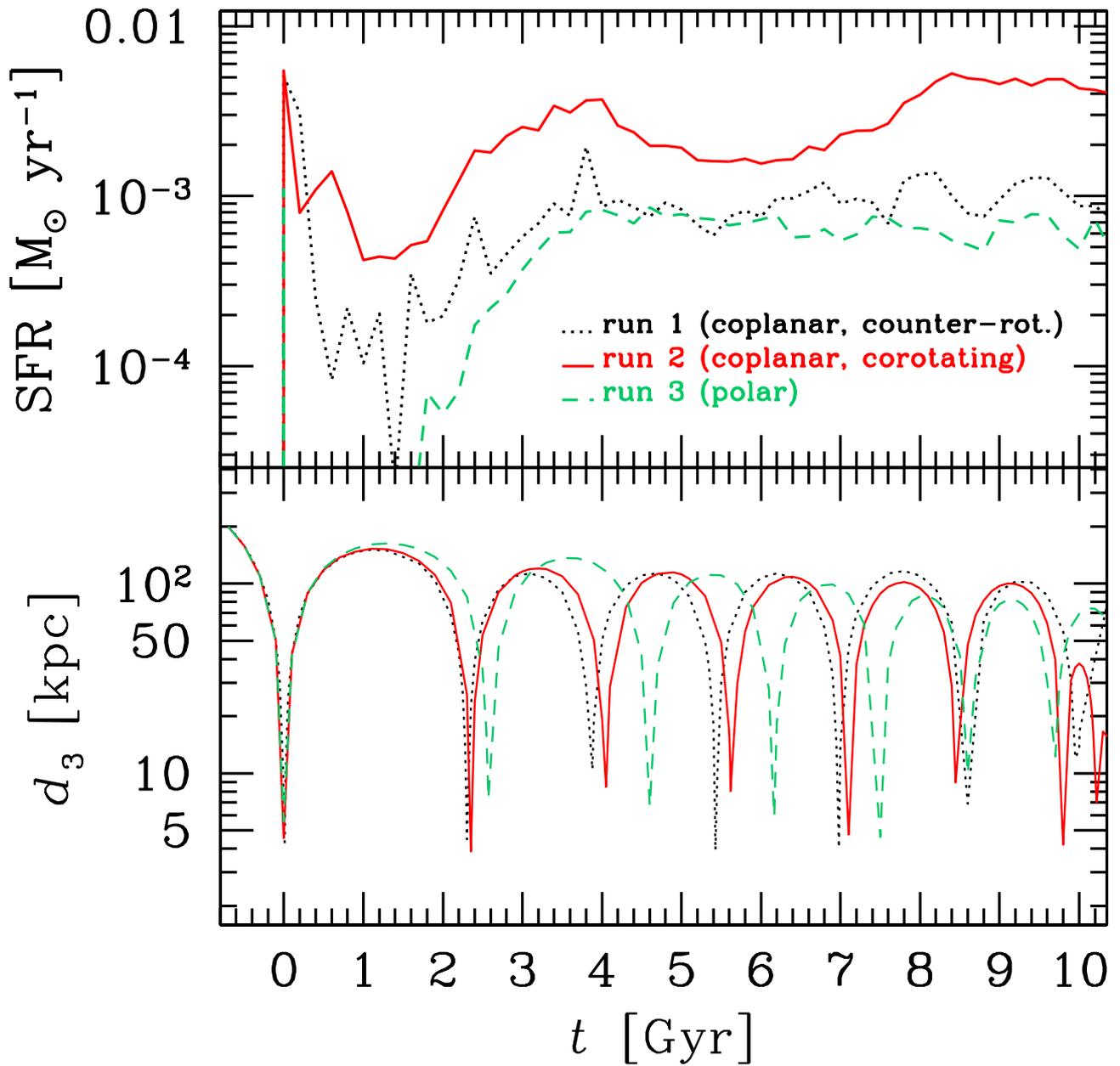,width=18.0cm} 
}}
\caption{\label{fig:fig2}
Top panel: SFR as a function of time in runs~1 (dotted black line), 2 (solid red line) and 3 (dashed green line). Bottom panel: three-dimensional distance between the centre of mass of the dwarf galaxy and that of the S0 galaxy, as a function of time, in the same runs. Time $t=0$ is the epoch of the first periapsis passage.
}
\end{figure*}

The bottom panel of Fig.~\ref{fig:fig2} shows the orbit of runs~1, 2 and 3 as a function of time. The time evolution of the orbit is initially the same in the three runs (because they have the same energy, angular momentum and impact parameter), but differences arise starting from the second periapsis passage, because dynamical friction and tidal dissipation affect each run differently. In each simulation, the dwarf galaxy undergoes seven periapsis passages in 10 Gyr.  At the end of the simulation, most of the stars that initially belonged to the dwarf galaxy reside in tidal streams and shells surrounding the S0 galaxy. They end up forming a stellar halo with distance $>50$ kpc from the centre of the S0 galaxy (see Figs. 6 and 7 of \cite{Mapelli2015}).

The top panel of Fig.~\ref{fig:fig2} shows the star formation rate (SFR) as a function of time. Before the first periapsis passage no star formation takes place in the dwarf galaxy (the only simulated galaxy that hosts gas). Immediately after the first periapsis passage, an instantaneous burst of star formation occurs in all simulations. This indicates that star formation is entirely triggered by the interaction. The burst of star formation that takes place after the first periapsis passage occurs in the dwarf galaxy (see \cite{Mapelli2012}), while no star formation occurs in the S0 galaxy, which has not yet accreted gas.

A second episode of star formation starts during the second periapsis passage, and lasts for the entire simulation. This second episode of star formation takes place at the centre of the S0 galaxy, which has accreted gas from the stripped dwarf galaxy. The secondary burst is stronger in run~2 (i.e. the coplanar corotating run). The SFR is a few $\times{}10^{-3}$ M$_\odot$ yr$^{-1}$ for the entire simulation in all runs, indicating a low level of star formation activity for a very long time span. 

This result indicates that minor mergers are able to re-activate star formation in an S0 galaxy for a very long time after the first periapsis passage. Our simulated SF rate is quite lower than the SF rate of rejuvenated S0 galaxies \cite{Salim2014}. This difference can arise from the fact that our simulated S0
galaxies are completely gas-free before the merger, while real S0 galaxies have a gas reservoir, in which SF can be triggered during the merger. We are currently investigating this issue in a follow-up study (Mazzarini et al., in preparation). Also, we note that most SF occurs in the nucleus of the S0 galaxy, rather than in the ring. This is likely a limitation of the stochastic recipes for SF (see \cite{Mapelli2015} for details); these cannot resolve SF in relatively diffuse warm gas regions (the ring), while enhance SF in the density peaks (the nucleus). 

\subsection{Gas rings}
\begin{figure}
\center{{
\epsfig{figure=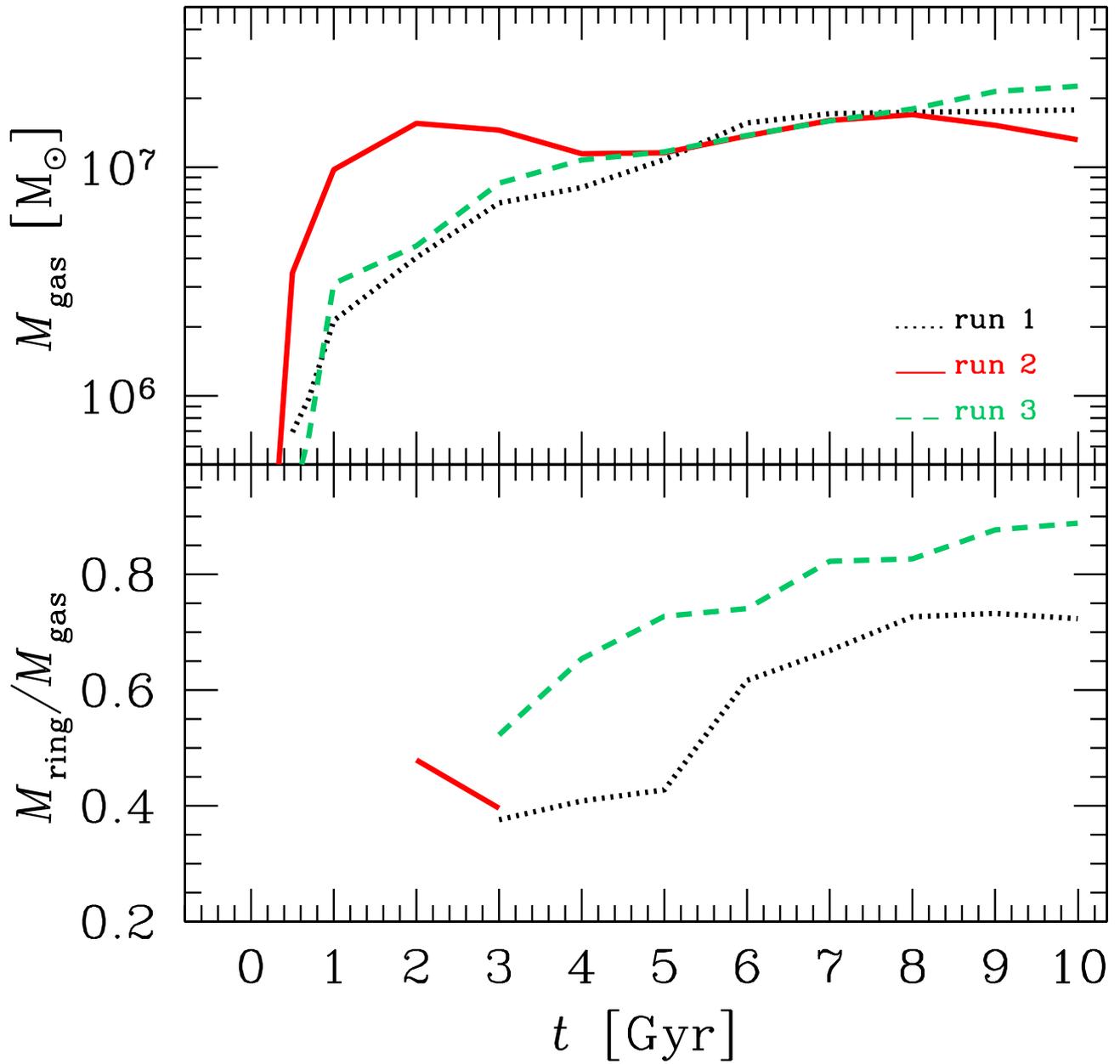,width=18.0cm} 
}}
\caption{\label{fig:fig3}
Top panel: gas mass $M_{\rm gas}$ in the innermost 15 kpc of the S0 galaxy, as a function of time. Bottom panel: fraction of gas mass that lies in a cold ring ($M_{\rm ring}/M_{\rm gas}$), as a function of time. Dotted black line: run~1, solid red line: run~2; dashed green line: run~3.}
\end{figure}

\begin{figure*}
\center{{
\epsfig{figure=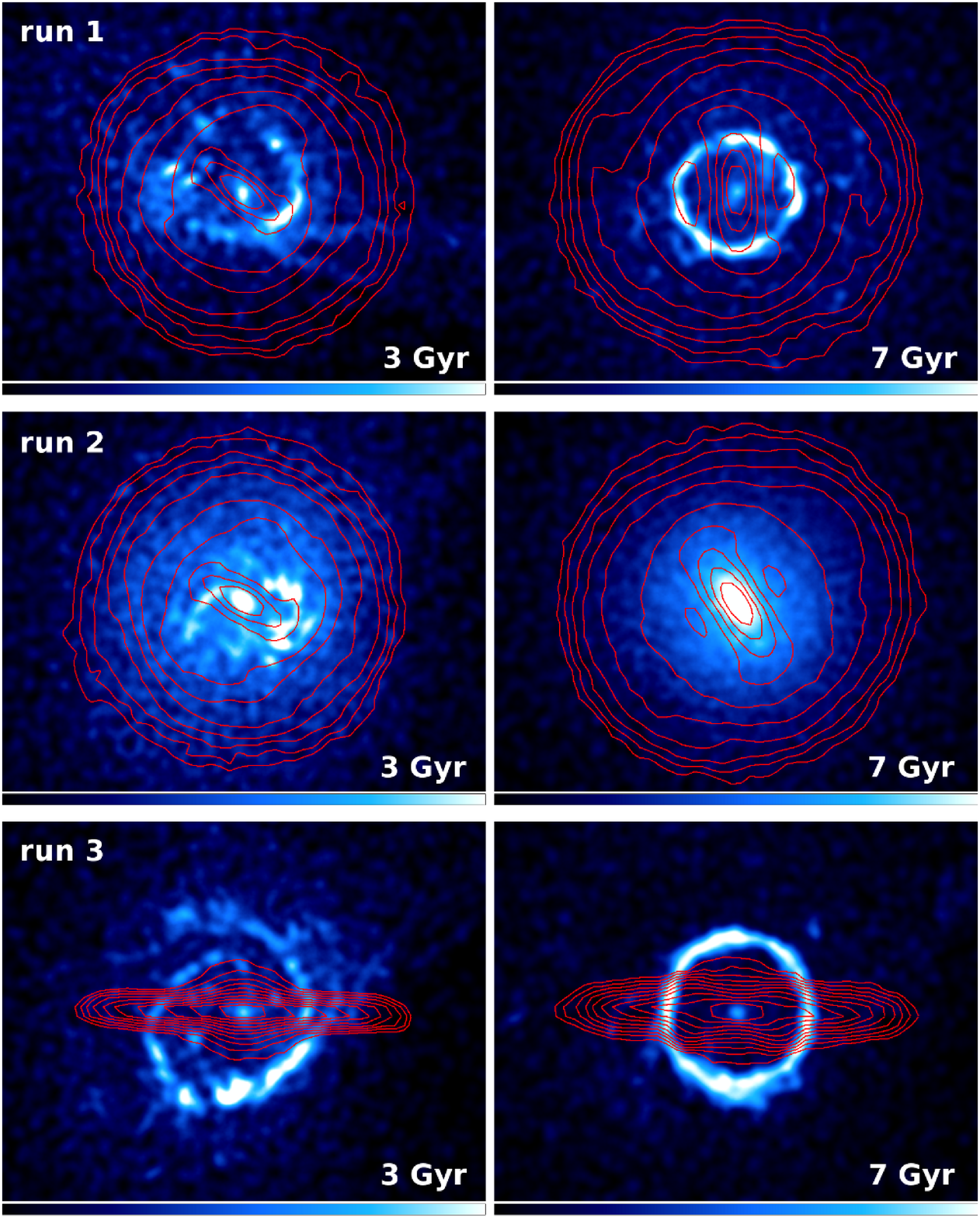,width=18.0cm} 
}}
\caption{\label{fig:fig4}
Projected density of gas (logarithmic colour-coded map) and stars (red isocontours) of the primary galaxy. The gas is projected face on. Top panels: run~1 at 3 Gyr (left) and 7 Gyr (right). Central panels: run~2 at 3 Gyr (left) and 7 Gyr (right). Bottom panels: run~3 at 3 Gyr (left) and 7 Gyr (right). The density of gas particles was smoothed with a cubic spline. Each box measures $41\times{}32$ kpc.
}
\end{figure*}

About one tenth of the total initial gas mass of the dwarf galaxy ends up in the innermost 15 kpc of the S0 galaxy \cite{Mapelli2015}. 
A fraction of this gas forms a ring-like structure in the S0 galaxy. Fig.~\ref{fig:fig3} shows the fraction of gas mass that lies in a warm ($\sim{}10^4$ K) gas ring, with a radius of $\sim{}5-15$ kpc (depending on the simulation). In runs~1 and 3, the ring forms at $\sim{}3$ Gyr and lasts for the entire simulation. In run~2, the ring ring forms at $\sim{}2$ Gyr, but is already destroyed at $\sim{}3$ Gyr (and gas from the ring is funnelled to the nucleus of the S0 galaxy).

Therefore, a gas ring forms in all simulations, but it is short-lived in the corotating case (run~2). If the orbit of the satellite galaxy is either polar or coplanar and counter-rotating, the ring forms later but is long-lived and fairly stable. Fig.~\ref{fig:fig4} shows the map of gas in runs~1, 2 and 3 at time $t=3$ and 7 Gyr after the first periapsis passage. At $t=3$ Gyr, a ring is present in all runs, but it is quite clumpy and perturbed. At $t=7$ Gyr, the ring has become regular and smooth in runs~1 and 3, while it has completely disappeared in run~2. In run~3, the ring is polar, thus preserving the initial orbital inclination of the satellite galaxy. In runs~1 and 2, the ring counter-rotates and corotates with the stars of the S0 galaxy, respectively, preserving the initial orbital angular momentum direction of the satellite galaxy.

 We note that the dark matter halo in our simulations is spherical, and is not significantly perturbed during the merger. Whether polar rings can help us constraining the shape of dark matter halos has been debated for a long time (e.g. \cite{Combes2013} and references therein). Theoretical work indicates that triaxial halos can support  polar rings, if their tumble period is long with respect to the orbital period of the polar ring (e.g. \cite{Steiman-Cameron1982}). On longer timescales, the polar ring can become warped or unstable. Moreover, some observations \cite{Iodice2003} suggest that the velocity field of some polar ring galaxies is consistent with a flattened dark matter halo, along the plane of the polar ring. Our simulations show that stable polar rings form also if the dark matter halo is spherical.

These results indicate that smooth, regular and long-lived rings can form from minor mergers, if the initial satellite orbit was counter-rotating or polar with respect to the plane of the disc of the S0 galaxy.

\section{Conclusions}

In this  proceeding, we discussed the importance of minor mergers to rejuvenate S0 galaxies. We studied the merger of a gas-rich dwarf galaxy with an S0 galaxy, by means of N-body/SPH simulations, considering different orbital parameters of the dwarf galaxy. We found that minor mergers can re-activate star formation (with SFR $\sim{}10^{-3}$ M$_\odot$ yr$^{-1}$) in S0 galaxies for a long time ($\approx{}10$ Gyr). 

A warm ring of gas forms in the S0 galaxy, as a consequence of the tidal disruption of the dwarf galaxy. The ring is initially clumpy and irregular in all runs. If the orbit of the satellite galaxy was corotating with the disc of the S0 galaxy, the ring is short-lived ($<1$ Gyr). In contrast, if the orbit of the satellite galaxy was polar or counter-rotating with respect to the disc of the S0 galaxy, the ring is long-lived and becomes increasingly smooth and regular. This effect is probably driven by bar resonance. The ring preserves the initial orbital properties of the satellite galaxy: it is corotating (counter-rotating) with the disc of the S0 galaxy, if the orbit of the satellite galaxy was corotating (counter-rotating). Mergers with a satellite galaxy on a polar orbit produce long-lived polar rings. Thus, minor mergers can account for a (large) fraction of star forming S0 galaxies, since they trigger both a long-lived episode of star formation and the formation of a smooth long-lived gas ring.


\acknowledgments{Acknowledgments}
MM thanks the organizers of the EWASS-2015 conference, the organizers and the participants of the Special Session 3 (3D view on interacting and post-interacting galaxies from clusters to voids) for the excellent conference, the stimulating discussions, and the terrific time spent in Tenerife. This research is supported by INAF through PRIN-2014-14.





\conflictofinterests{Conflicts of Interest}

The authors declare no conflict of interest.

\bibliographystyle{mdpi}
\makeatletter
\renewcommand\@biblabel[1]{#1. }
\makeatother


%

\end{document}